\documentclass[english,aip,apl,reprint]{revtex4-1}
\usepackage[T1]{fontenc}
\usepackage[latin9]{luainputenc}
\usepackage{amsbsy}
\usepackage{amstext}
\usepackage{amssymb}
\usepackage{graphicx}
\usepackage{babel}
\begin{document}
\title{Efficient Identifying the Orientation of Single NV Centers in Diamond
and Using them to Detect Near Field Microwave}
\author{Xuerui Song}
\email{songxr@cast504.com}

\author{Fupan Feng}
\email{fengfp@cast504.com}

\author{Chunxiao Cai}
\affiliation{China Academy of Space Technology (Xi'an), Xi'an, Shaanxi, 710100,
P. R. China}
\author{Guanzhong Wang}
\author{Wei Zhu}
\affiliation{Hefei National Laboratory for Physical Science at Microscale, and
Department of Physics, University of Science and Technology of China,
Hefei, Anhui, 230026, P. R. China}
\author{Wenting Diao}
\author{Chongdi Duan}
\affiliation{China Academy of Space Technology (Xi'an), Xi'an, Shaanxi, 710100,
P. R. China}
\begin{abstract}
Arrays of NV centers in the diamond have the potential in the fields
of chip-scale quantum information processing and nanoscale quantum
sensing. However, determining their orientations one by one is resource
intensive and time consuming. Here, in this paper, by combining scanning
confocal fluorescence images and optical detected magnetic resonance,
we realized a method of identifying single NV centers with the same
orientation, which is practicable and high efficiency. In the proof
of principle experiment, five single NV centers with the same orientation
in a NV center array were identified. After that, using the five single
NV centers, microwave near field generated by a $20\,\mu m$-diameter
Cu antenna was also measured by reading the fluourescence intensity
change and Rabi frequency at different microwave source power. The
gradient of near field microwave at sub-microscale can be resoluted
by using arry of NV centers in our work. This work promotes the quantum
sensing using arrays of NV centers.
\end{abstract}
\maketitle
Nitrogen-vacancy(NV) centers are paramagnetic defects within the diamond
lattice, which consist of a substitutional N imputiry adjacent to
a vacant lattice site. Owing to their outstanding optical and spin
properties at room temperature, NV centers have attracted more and
more attentions in the fields of nanoscale quantum sensing and quantum
information processing.\cite{degen2017quantum,maze2008nanoscale,ariyaratne2018nanoscale,maurer2012roomtemperature,bradley2019atenqubit,zaiser2016enhancing}
Especially in the fields of quantum sensing, due to the fact that
the NV center can be stablely located colse to the diamond surface
at nanoscale or nanodiamonds,\cite{osipov2019photoluminescence,song2013generation,maletinsky2012arobust,ofori-okai2012spinproperties}
the spatial resolution of the sensors based on NV centers have down
to nanoscale.\cite{hall2009sensing,ariyaratne2018nanoscale,kucsko2013nanometrescale,schmitt2017submillihertz}
This nanoscale property is quite important in the fields of condense
matter physics and biology research.\cite{casola2018probing,wu2016diamond,shi2018singledna,lillie2019imaging}
Recently, some groups reported that the near field microwave(MW) sensing
using NV centers in diamond have down to micro even nanoscale resolutions.\cite{wang2015highresolution,appel2015nanoscale,shao2016widefield,dong2018afiber,horsley2018microwave,mariani2018imaging,chipaux2015widebandwidth} 

Precise detecting of the spatial distribution of the MW near field
is crucial for developing new types of MW devices, chip failure checking,
electromagnetic compatibility analyzing and even solid state physics
studying.\cite{wallraff2004strongcoupling} Generally speaking, the
spatial resolution of the traditional dipole probe is limited to about
$100\,\mu m$. Even worse, during the MW measuring process, the dipole
probe will reconstruct the electromagnetic field, which can disturb
the near field detection.\cite{holloway2014broadband} In recent
years, new sensors based on alkali vapor cell, superconductor and
spin-torque diode have been developed to detect the near field MW
with high sensitivity.\cite{fang2016giantspintorque,couedo2019hightcsuperconducting,fan2015effectof}
However, none of them obtained nanoscale spatial resolution. The development
of near field MW sensors based on NV center will bring an ultimate
solution to this problem.

As shown in the right section of Fig. 1, the ground state of electron
spin of the NV center is a spin triplet $^{3}A_{2}$ with a zero field
splitting of $D=2.87\,GHz$. The oscillation between $\vert0\rangle$
and $\vert\pm1\rangle$ ground states can be driven by a resonant
MW magnetic field as $\hbar\omega=D\pm\gamma B_{Z}$, where $B_{Z}$
is a static magnetic field along the axis of NV center, $\gamma=2.8025\,MHz/Gauss$
is the gyromagnetic ratio of the electron spin. The MW magnetic field
is written as $\boldsymbol{B}_{MW}(t)=\boldsymbol{B}_{MW}cos(\omega t)$,
where $f=\frac{\omega}{2\pi}$ is the MW frequency. By carrying out
the optical detected magnetic resonant (ODMR), the resonant frequency
of the MW field can be obtained. By adding static magnetic field $\boldsymbol{B}$,
$\vert\pm1\rangle$ can be further splitted, which varies the resonant
MW field frequency from kHz to sub-THz.\cite{aslam2015singlespin}
Additionally, the Rabi frequency of the spin state oscillation $\Omega$
is proportional to the component of the MW field in a plane perpendicular
to the NV axis as $\Omega=\gamma B_{MW\perp}$. At last, by using
NV centres at the four different axes, the total amplitude $B_{MW}$
and the orientation of the vector can be obtained by mearsuring different
$B_{MW\bot}$. By doing this, nonivasive detecting of near field MW
magnetic field at nanoscale resolution obtained.\cite{wang2015highresolution,appel2015nanoscale}

Traditionally, to measure field at nanoscale resolution, scanning
should be carried out by combining the NV sensors with AFM tip, which
is highly requirement in practice.\cite{appel2015nanoscale,maletinsky2012arobust,degen2008scanning}
Alternatively, by measuring field using NV centers at different sites,
NV center arrays in the diamond have the potential in the fields nanoscale
quantum sensing.\cite{momenzadeh2015nanoengineered,wang2015highsensitivity,fukami2019alloptical}
Howerer, determining their orientations one by one is resource intensive
and time consuming. In this work, using shallow NV center arrays in
diamond, a method by combining scanning confocal fluorescence image
and optical detected magnetic resonance was developed to efficiently
identify the NV centers with same orientation. The relative MW amplitude
distribution is obtained by measuring the fluorescence intensity and
the Rabi oscillation of the single NV center at different sites of
the arrays. The spatial resolution of near field MW sensors based
on array of NV centers in this work is hundreds of nanometers which
consists with the optical microscope. 

\begin{figure}
\includegraphics[width=8cm]{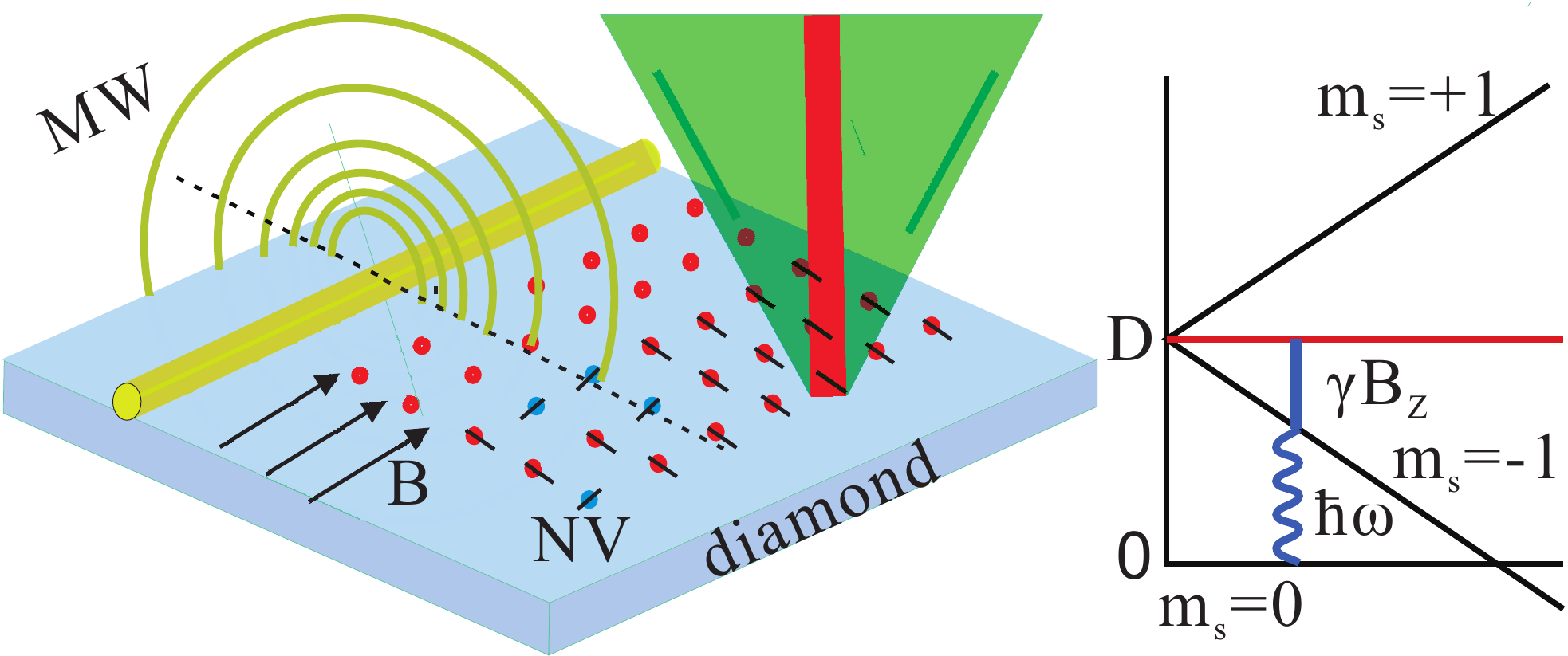}\caption{Experiments and principle illustrating of the work. Left, shallow
array of NV centers with different orientations were located close
to the diamond surface within nanometers. The near field MW was generated
by a $20\,\mu m$-diameter Cu antenna which lays tens of micrometers
above the diamond surface. The NV center was exciated by $532\,nm$
green laser. The red fluorescence was collected by an objective and
then detected by an APD. Right, energy level of the ground state of
the electron spin of NV center.}
\end{figure}

The sample used in this work was a $2\times2\times0.5\,mm^{3}$ high-quality
electronic grade diamond with natural isotopic concentration of $^{13}C$
(1.1\%) from Element Six. The NV center array was prepared by implanting
$8\text{\,}keV$ $^{14}N_{2}^{+}$ molecules with a fluence of $8\times10^{10}\,{}^{14}N_{2}^{+}/cm^{2}$.
The implantation angle is $7^{\circ}$ through $60\,nm$ diameter
apertures which were patterned using electron beam lithography in
a 300-nm-thick polymethyl methacrylate (PMMA) layer deposited on the
diamond surface.\cite{feng2016efficient,wang2015highsensitivity,toyli2010chipscale}
After implantation, the sample was annealed $2h$ at $1050\,^{\circ}C$
in a vacuum at $2\times10^{-5}\,Pa$ to induce vacancy diffusion to
form NV centers. According to SRIM simulation of the stopping of nitrogen
ions in diamond chip, the most probable stopping position of the implanted
nitrogen ions was $6.5\,nm$ below the diamond surface.\cite{ziegler2010srimtextendash}
After paperation of the NV center arrays, the diamond sample was placed
in the set-up shown in Fig.1. The scanning confocal fluorescence image
of the NV center array with a $10\times10\,\mu m^{2}$ region was
shown in Fig. 2(a). The fluence of the $^{14}N_{2}^{+}$ molecules
and the size of the apertures used during the implantation ensure
that most of the bright pots in the arrays was single NV center, which
consistents with the ODMR experiments. It should be noted that due
to the low enengy of the implanting ions, creation of the NV centers
were unsuccessful in some sites.

\begin{figure}
\includegraphics[width=8cm]{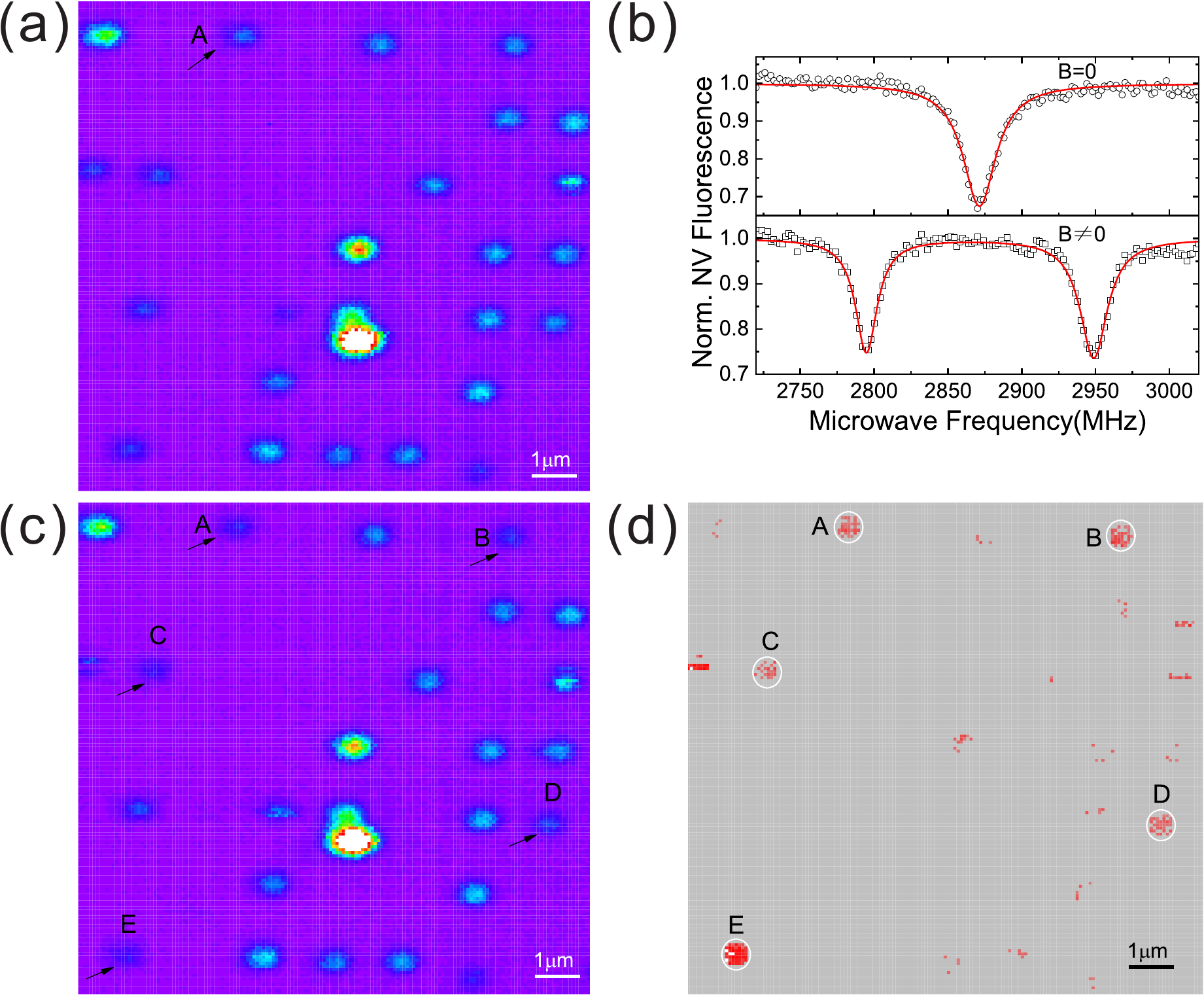}\caption{Efficient identifying the orientation of NV centres in diamond.(a)
Scanning confocal fluorescence image of NV center array with a $10\times10\,\mu m^{2}$
region. (b) ODMR spectrum of the NV center as pointed in Fig. 2(a).
Upper spectrum, $B=0$. Lower spectrum, $B\protect\neq0$. (c) Under
continues MW exciation with resonance frequency at $2.794\,GHz$ ($2.949\,GHz$),
the scanning confocal fluorescence image of the region of Fig. 2(a)
was recorded again. (d) Imaging of the five single NV centers with
the same orientation.}
\end{figure}

Then, by combining scanning confocal fluorescence image and optical
detected magnetic resonance, we realized a method which can fast identify
single NV centers with the same orientation, so called scanning optical
detected magnetic resonance (SODMR) method. By sweeping the MW frequency
using the $20\,\mu m$-diameter Cu antenna under continuous $532\,nm$
green laser exciation, the ODMR of a single NV center that randomly
selected ( pointed by an black arrow in Fig. 2(a)) under zero magnetic
field was measured. The dip of the ODMR spectrum (Fig. 2(b)) located
at $2.87\,GHz$ corresponds to the transition between $\vert0\rangle$
and $\vert\pm1\rangle$ states. After that, a static magnetic field
$\boldsymbol{B}$ generated by an electromagnet was applied to the
NV center to further split the $\vert\pm1\rangle$ states. At this
condition, as shown in lower spectrum of Fig. 2(b), the two dips located
at $2.794\,GHz$ and $2.949\,GHz$ correspond to the transitons of
$\vert0\rangle\leftrightarrow\vert-1\rangle$ and $\vert0\rangle\leftrightarrow\vert+1\rangle$
respectively. As the resonance frequency of the A-point is known,
the NV centers with the same orientation can be identified immediately
by our SODMR method introduced here. By applying continues MW exciation
at $2.794\,GHz$ ($2.949\,GHz$), the scanning confocal fluorescence
image of the same region of Fig. 2(a) was recorded again. The dramatically
decreased fluorescence intensity of the A-point in Fig. 2(c) confirmed
that magnetic resonance occured during scanning counting of the fluorescence
of A. As same as A, the fluorescence intensity of the other four single
NV centers marked as B,C,D and E, which have the same orientation
as A, also decreased. As a comparison, the fluorescence intensity
of the other NV centers, which have  different orientation with A,
showed litte change. To further improve the efficiency of identifying
of the NV centers with the same orientation, the contrast diagram
of the fluorescence intensity was normalization as $\frac{I_{res}-I_{no-MW}}{I_{no-MW}}$,
where $I_{res}$ (Fig. 2(c)) was the fluorescence of NV centers under
microwave radiation and $I_{no-MW}$ (Fig. 2(a)) was that of without
MW. As shown in Fig. 2(d), five of the single NV centers with the
same orientation obviously displayed. The method of identifying of
the single NV centers with the same orientation introduced here is
more efficiency for arrays with a mass of NV centers.\cite{toyli2010chipscale,hausmann2011singlecolor}

\begin{figure}
\includegraphics[width=8cm]{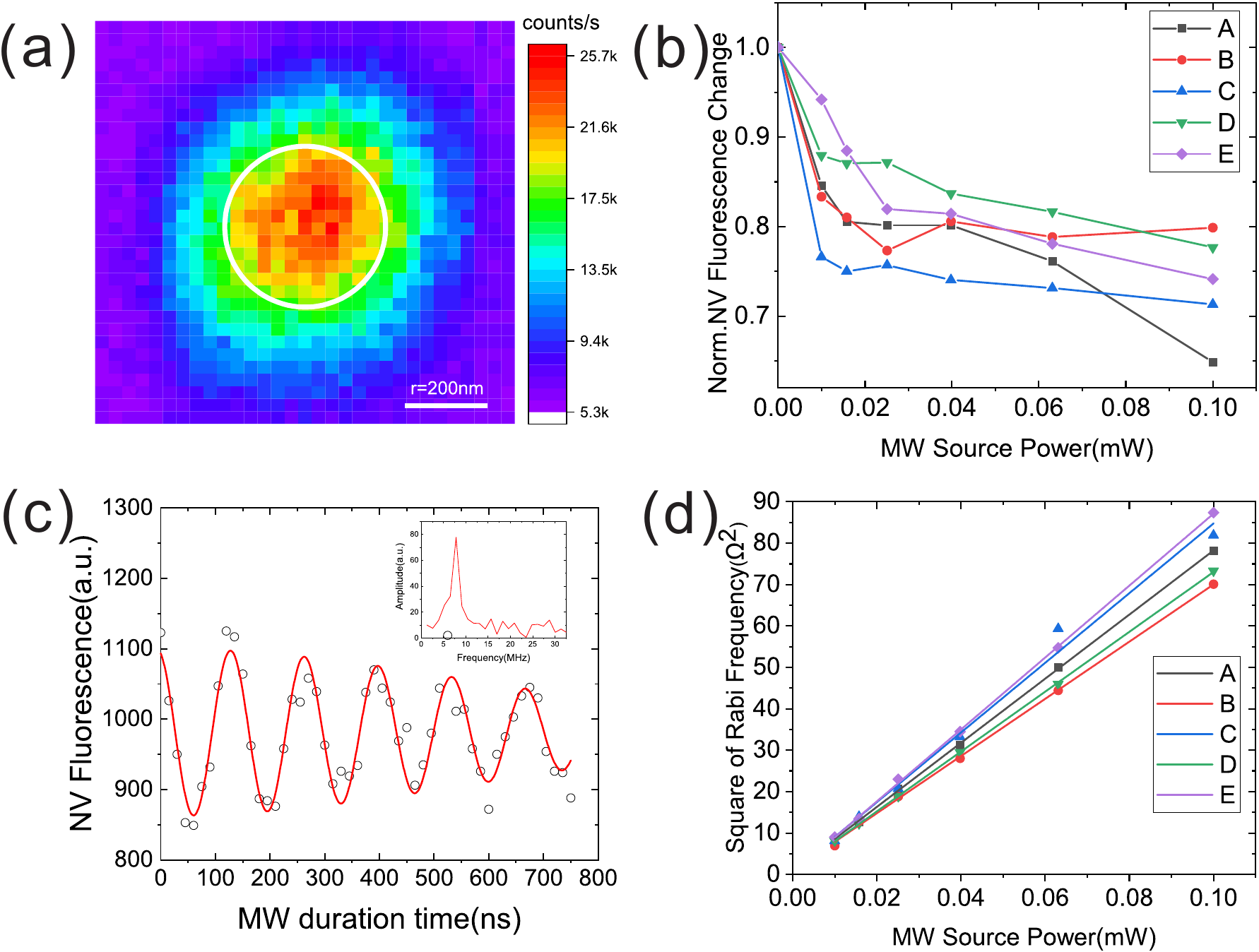}\caption{Using the five NV centers identified, MW near field generated by a
$20\,\mu m$-diameter Cu antenna was measured. (a) Scanning confocal
fluorescence image of one of the single NV center at MW power of $0.04\,mW$.
(b) NV fluorescence change VS MW source power.(c) Example of Rabi
oscillation of single NV center. Insert, FFT of the Rabi oscillation.(d)
Square of Rabi frequency($\Omega^{2}$) versus MW source power.}

\end{figure}

As has been reported, the MW power influences the contrast of the
ODMR spectrum.\cite{jensen2013lightnarrowing,dreau2011avoiding}
Therefore, on the condition of resonance, the power of MW near field
can be indirectly read out by measuring the fluorescence intensity
change of the NV centers.\cite{appel2015nanoscale,yang2019usingdiamond}
Utilizing the five NV centers with the same orientation identified
above, the relative component of the MW field amplitude in a plane
perpendicular to the NV axis $B_{MW\perp}$ was studied by measuring
the fluorescence intensity and Rabi oscillation frequency $\Omega$.
Fig. 3(a) shows the scanning confocal fluorescence image of single
NV center at MW power of $0.04\,mW$. The corresponding fluorescence
intensity was obtained by sum the total counts within a circle around
the center of fluorescence spot with a radius of $200\,nm$. Fig.
3(c) shows an example of the Rabi oscillation, and the Rabi frequency
was obtained by carrying out the FFT of the data(insert in Fig. 3(c).
The measured projected fluorescence intensity and square of Rabi frequency($\Omega^{2}$)
versus MW source power were shown in Fig. 3(b) and Fig. 3(d) respectively.
From Fig. 3(b) we can see that, with the increasing of the MW source
power from $0$ to $0.02\,mW$, the NV fluourescence intensity sharply
decreased. From $0.02$ to $0.1\text{\,}mW$, the fluourescence intensity
drecreased steadly, indicating that a saturation effect exists under
strong MW source power.\cite{dreau2011avoiding} This result demonstrated
that the method of measuring the relative MW intensity by reading
the NV fluourescence intensity change is more faithful under weaker
MW power excitation. Additionally, under strong MW excitation(e.g.
$0.1\,mW$), the NV fluourescence intensity change of A,C,E is more
obvious than B and D, which consistents with the fact that the later
two NV centers are more far-away from the Cu MW antenna, as the configuration
shown in Fig. 1 and Fig. 2(c). Fig. 3(d) shows the proportional relationship
of the square of Rabi frequency ($\Omega^{2}$) versus MW source power
as $\Omega^{2}\propto P$. The reslut datas were fitted by linearity
curves. The gradient of the MW was precisely reflected by the slopes
of the linearity curves as $E>C>A>D>B$, which is consistent with
the configuration of the five NV centers shown in Fig. 2(c), except
for the spot of B. The unusual of spot B was also shown in the Fig.
3(b) under strong MW power(e.g. $0.1\,mW$), which indicated the distribution
of MW near field is not simply linearity with the distance of the
antenna.\cite{appel2015nanoscale} The result indicated that the
gradient of near field MW at sub-microscale can be resoluted by using
arry of NV centers in our work. In this paper, only the relative MW
amplitude component perpendicular to the quantum axis of the five
NV centers with the same orientation was obtained. The MW amplitude
vector can be reconstructed by further measuring the Rabi frequency
or fluourescence intensity change by using the NV centers in other
three axis.\cite{wang2015highresolution} In this work, near field
MW was detected by reading the fluourescence intensity change and
Rabi frequency. Compared with the method by reading the fluourescence
intensity change, the method by detecting the Rabi frequency is more
convincing to refelct the distribution of the near field MW at high
power.

In conclusion, by combining scanning confocal fluorescence images
and optical detected magnetic resonance, we realized a method which
can fast identify single NV centers with the same orientation. By
using an array of single NV center in diamond, five single NV centers
with the same axis were identified in a $10\times10\,\mu m^{2}$ region
which demonstrated that the method is practicable and high efficiency.
After that, the relative component of the MW field amplitude in a
plane perpendicular to the NV axis was studied by measuring the fluorescence
intensity and Rabi oscillation frequency. The gradient of near field
MW at sub-microscale can be resoluted by using arry of NV centers
in our work. The method of identifying the orientation of single NV
centres in diamond introduced here can also be used to classify spins
with different orientations in other solid crystals such as SiC.\cite{wang2017efficient}
By designing the unique configuration, the near field MW sensor using
the NV center array introduced here have the potential in the field
of chip failure checking.
\begin{acknowledgments}
This work was supported by the Natural Science Foundation of China
(Grant No. 11505135 and No. 61501368) and the National Key Laboratory
Foundation (Grant No. 6142411185307).
\end{acknowledgments}

\bibliographystyle{bibtex/bst/revtex/aipnum4-1}

\end{document}